\documentclass[10pt,a4paper]{article}

\usepackage{amsfonts}
\usepackage{amssymb}
\usepackage{amsmath}
\usepackage{graphicx,subfigure}

\usepackage{pgf,tikz}
\usetikzlibrary{arrows}
\usetikzlibrary{snakes}
\usetikzlibrary{shapes.arrows}
\usetikzlibrary{matrix}
\usepackage{bm}
\usepackage{float}

\newcommand{\ket}[1]{|{#1}\rangle}
\newcommand{\bra}[1]{\langle{#1}|}

\newcommand{\exval}[1]{\langle{#1}\rangle}

\usepackage{bbm}
\usepackage{dsfont}

\begin{document}
\title{The rhythm of quantum algorithms\footnote{The paper is going to appear on \textit{Soft Computing}.}}

\author{Stefano Bonzio\thanks{Department of Pedagogy, Psychology and 
Philosophy, University of Cagliari. E-mail:stefano.bonzio@gmail.com}\and 
Paola Verrucchi\thanks{Istituto dei Sistemi Complessi, ISC-CNR; 
Dipartimento di Fisica ed Astronomia dell'Universit\`a di Firenze; INFN 
Sezione di Firenze; Via G.Sansone 1, 50019 Sesto Fiorentino (FI), Italy. 
E-mail: verrucchi@fi.infn.it}} \date{} \maketitle \thispagestyle{empty}

\begin{abstract}
Quantum algorithms can be generally represented as the dynamical 
evolution of an input quantum register, with the action of each logical 
gate, as well as of any transmission channel, defined by some
quantum propagator. From a global viewpoint this unitary dynamics is 
ruled by the flow of a continuous time, and the possible splitting
into shorter logical sub-units is nothing but a harmless, though 
useful, zooming process. On the other hand, understanding how
elementary units of the quantum register, namely single qubits, are 
actually hauled along the algorithm, is a more complex matter, 
as it involves the dynamical entanglement generation entailed in 
the action of two-qubit gates.
In this work we first review how the essential elements of 
quantum algorithms can be described in terms of dynamical processes, and 
then analyze the corresponding non-unitary dynamics of single qubits, by 
referring to the formalism adopted in the study of open quantum systems.
We show that single qubits evolution cannot be split into intervals 
shorter then the typical time needed by two-qubit gates for 
accomplishing their task, which somehow gives a rhythmical structure to 
the algorithm itself.
We further point out that the local evolution entails a memory, in that 
the way each qubit takes an infinitesimally small step forward in 
time, is set by its previous history, back to the instant when it 
entered the last two-qubit gate. This memory originates from quantum 
correlations, and it is suggested to play an essential role in quantum 
information processing. As a concluding remark we just touch on the idea 
that a similar analysis could be put forward for getting a clue on how 
we extract meaningful contents out of complex informational input.

\end{abstract}
\section{Introduction}

The idea that computation be a dynamical process is quite natural, 
either when referring to the classical or to the quantum case:
the notion stems from the very way we face a calculation, step 
after step towards the final result, but it can actually imply a 
completely different role of the fundamental parameter of any dynamical 
process, namely "time".

In classical computation it is not usually necessary to refer to a 
physical time, in that a discrete index labelling the subsequent steps 
of the algorithm is sufficient to describe its functioning.
Although this keeps holding when analyzing quantum algorithms 
from a purely logical viewpoint, their description as
physical processes, ruled by the principles of quantum mechanics, 
requires that time be the actual, continuous, parameter governing any
dynamical evolution.

There is no apparent conflict between the above pictures, based on 
discrete steps and continuous time, respectively:
after all the former can be indentified as finite intervals of the latter, 
and the latter as the limit for increasingly shorter 
intervals. However, this limit cannot always be taken: aim of this 
article is in fact making this point clearer referring to some 
fundamental aspects of the dynamics of open quantum systems (OQS) and 
their role in making quantum algorithms essentially different from 
classical ones.

In Sec.\ref{s.elements} we briefly review the essential elements of any 
quantum device, and describe their dynamical functioning from a physical 
viewpoint. In Sec.\ref{s.entanglement} we specifically consider 
entangling gates \cite{DiVi95} and introduce the formalism of dynamical 
maps into the analysis. This allows us to
underline how non-markovianity emerges as a 
key-feature of entangling dynamical processes, as discussed in 
Sec.\ref{s.non-markovian}. 
Inspired by the relation between markovianity and memorylessness we
comment upon the role that memory can play in information processing and, in 
Sec.\ref{s.memory} we finally suggest possible implications of the 
presented analysis in more general contexts, such as those regarding the 
way we elaborate complex musical events.

\section{Quantum algorithms as dynamical processes}
\label{s.elements}
The grafical representation of quantum algorithms helps visualizing 
their main components and essential features. \\
\begin{figure}[H]
\begin{center}
\begin{tikzpicture}
\draw (-10,1.5) node {$\ket{Q_1}_{\rm in}$};
\draw (-10,-1.5) node {$\ket{Q_2}_{\rm in}$};
\draw (-7,1.8) node {(1)};
\draw (-9,1.5) --  (-5,1.5)[line width=1.5];
\draw (-9,-1.5) --  (-8,-1.5)[line width=1.5];
\draw (-8,-0.5) rectangle (-6,-2.5)[line width=1];
\draw (-6,-2.8) node {(2)};
\draw (-7,-1.5) node {\large G1};
\draw (-5,2.5) rectangle (-3,-2.5) [line width =1];
\draw (-3,-2.8) node {(3)};
\draw (-4,0) node {\large G2};
\draw (-6,-1.5) -- (-5,-1.5)[line width =1.5];
\draw (-3,-1.5) -- (-1,-1.5) [line width =1.5];
\draw (-3,1.5) -- (-2,1.5) [line width =1.5];
\draw (-2,2.2) rectangle (0,1) [line width =1];
\draw (0,.7) node {(4)};
\draw (-1,1.03) -- (-.2,1.65) [->,line width =1.5];
\draw (0,1) arc (0:180:1);
\draw (0,-1.5) node {$\ket{Q_2}_{\rm out}$};
\end{tikzpicture}
\end{center}
\caption{Graphical representation of a quantum algorithm: (1) quantum channel, (2) one-qubit logical gate, (3) two-qubit logical gate, (4) measurement apparatus.}
\label{f.qalgorithm}
\end{figure}
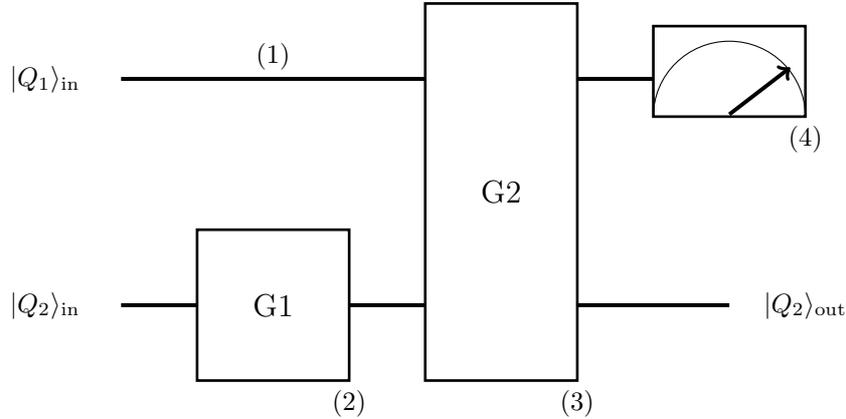
Referring to 
Fig.~\ref{f.qalgorithm}, besides the input/output 
quantum registers, one identifies quantum channels (1), single-qubit 
logical gates (2), two-qubit logical gates (3), and measurement 
apparatuses (4). Although they all represent dynamical processes, there 
is a substantial difference between (1)-(2) and (3)-(4), in that they 
correspond to unitary and non-unitary evolutions of single qubits, 
respectively. In order to better understand the meaning and 
the implications of this statement, in this section we briefly review 
the actual functioning of each element, starting from the primary
definition of the object of the processing, namely the qubit state.

A qubit is a physical system $Q$ whose Hilbert space ${\cal H}_Q$ has 
dimension dim${\cal H}_Q=2$. Once a {\it computational} basis
$\{\ket{0},\ket{1}\}$ has been chosen, 
and conventionally defined after $\sigma^z\ket{0}=-\ket{0}$ 
and $\sigma^z\ket{1}=\ket{1}$, with $\sigma^{\alpha}$ the Pauli matrices 
($\alpha=x,y,z$), a qubit state is
\begin{eqnarray}
&~&\ket{Q}=q_1\ket{1}+q_0\ket{0}~~~\rightarrow~~~
\left(
\begin{array}{cc}
q_1 \\
q_2 
\end{array}
\right)~,
\label{e.qubitstate}\\
&~&\exval{Q|Q}=|q_1|^2+|q_0|^2=1~,\label{e.qubitnorm}
\end{eqnarray}
where by $\rightarrow$ we will hereafter indicate the 
corresponding matrix-representation for vectors and operators.
The state evolution is described by
\begin{equation}
\ket{Q(t_2)}=U_{t_2,t_1}\ket{Q(t_1)}~~~{\rm with}~~~
U_{t_2,t_1} U^\dagger_{t_2,t_1}=\mathbbm{1}~~~{\rm for~any}~~~
t_1,t_2~,
\label{e.qubitevolution}
\end{equation}
where the unitarity of $U_{t_2,t_1}$ guarantees that the norm of the 
quantum state is conserved.
If time is homogeneous, as will be hereafter assumed, 
there always exists a Hermitian operator $H_{{Q}}$, acting on ${\cal 
H}_Q$ and constant in time, such that 
($\hbar=1$ throughout this paper)
\begin{equation}
\ket{Q(t)}=e^{-iH_{_{Q}}t}\ket{Q(0)}\equiv U_t^{(Q)}\ket{Q(0)}~~~{\rm 
with}~~~t\equiv
t_2-t_1~:
\label{e.HQ}
\end{equation}
this operator can be readily identified with the qubit Hamiltonian.

A quantum register is a system $QR$ made of $N>1$ qubits; its states 
$\ket{QR}$ are normalized elements of the Hilbert space ${\cal 
H}_{QR}=\otimes_i{\cal H}_{Q_i}$, that evolve according to
\begin{equation}
\ket{QR(t)}=e^{-iH_{_{QR}}t}\ket{QR(0)}\equiv U_t^{(QR)}\ket{QR(0)},
\label{e.QRevolution}
\end{equation}
where $H_{{QR}}$ is the global Hamiltonian, acting on ${\cal H}_{QR}$.
If $\ket{QR}$ is separable, $\ket{QR}=\otimes_i\ket{Q_i}$, each 
qubit is still described by the, usually dubbed {\it pure}, state 
$\ket{Q_i}$. If $\ket{QR}$ is entangled, 
$\ket{QR}\neq\otimes_i\ket{Q_i}$, there will be at least two qubits that 
need being described in terms of their respective , usually  dubbed 
{\it reduced}, density operator
\begin{equation}
\rho_{Q_i}\equiv{\rm Tr}_{{\cal H}_{QR}/{\cal H}_{Q_i}}\ket{QR}\bra{QR}
\rightarrow
\left(\begin{array}{cc}
\rho_{11} & \rho_{10} \\
\rho_{10}^* & \rho_{00} \\ \end{array}\right)~,
\label{e.rhoQi}
\end{equation}
where Tr$_{{\cal H}_{QR}/{\cal H}_{Q_i}}$ is the partial trace on 
the subspace ${\cal H}_{QR}/{\cal H}_{Q_i}$, and $\rho_{Q_i}$ 
consequently acts on ${\cal H}_{Q_i}$ only.
Due to the normalization of $\ket{QR}$ and the definition of the partial 
trace, $\rho_{Q_i}$ has the following properties (we drop the index 
$Q_i$ for the sake of a lighter notation):
\begin{equation}
i)~\rho^\dagger=\rho~~;~~ii)~{\rm Tr}\rho=1~~;~~iii)~{\rm Tr}\rho^2\leq 
1~~;~~{\it i)}~\exval{A|\rho|A}\ge 0~~{\rm for~any}~~\ket{A}\in{\cal H}~.
\label{e.rhoprop}
\end{equation}
The introduction of the reduced density operator allows one to formally 
define open quantum systems (OQS) as those whose behaviour can only 
be described in terms of $\rho$, due to their being entangled with some 
other quantum system, playing the role of their environment 
\cite{BreuerP02,Rivas12}.
The requirement that the 
properties \eqref{e.rhoprop} hold at all times, strongly characterizes 
the evolution of density operators, and hence the OQS dynamics.

\subsection{Quantum channels}
\label{s.quantumchannels}
A quantum channel, usually pictured as a single line, represents any 
process that realizes a point-to-point 
transfer of the quantum information embodied into the state of one 
single qubit $\ket{Q}$. Requiring that this state be the same at the 
left- and right-end of the channel, implies 
\begin{equation}
\ket{Q}_{\rm in}\equiv\ket{Q(t_1=0)}=
\ket{Q(t_2=\tau_{\rm ch})}e^{i\varphi}\equiv
\ket{Q}_{\rm out}~,
\label{e.channel}
\end{equation}
where $\varphi$ is an unessential phase, possibly depending on the actual 
time $\tau_{{\rm ch}}$ needed to complete the transfer process.

\begin{figure}[H]
		\begin{center}
		\begin{tikzpicture}
\draw (-1,0) -- (3,0)[line width=2, -];

\draw (-2,0) node {$\ket{Q}_{\rm in}$};
\draw (5,0) node {$\ket{Q}_{\rm out}=\ket{Q}_{\rm in}$};
\draw (1,-.7) node {$\tau_{{\rm ch}}$};
		\end{tikzpicture}
		\caption{Ideal quantum channel.}
		\end{center}
\end{figure}
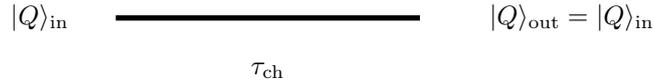
The above condition 
\eqref{e.channel} defines a perfect channel or, quite equivalently, a 
perfect quantum-state transfer process.
Depending on the specific physical realization of the circuit, such 
condition can be obtained in different ways; if the object embodying the 
qubit can physically move ("flying qubit"), than it is sufficient that 
its repositioning be shielded enough to avoid interactions other than 
those exclusively implying a global phase shift (this is for 
instance the case of qubits embodied in optical photons, moved around 
through optical fibers). On the other hand, if the qubit is encoded into 
the state of a particle that cannot move ("still qubit"), realizing a 
perfect quantum channel requires using an auxiliary medium, tipically a 
many-body system such as a spin-chain or an arrow of cold atoms in an 
optical lattice, with the interactions
properly designed so as to guarantee that condition \eqref{e.channel} be 
fullfilled\cite{BanchiV11}.
The time $\tau_{{\rm ch}}$, that depends on the relevant couplings 
and the physical length of the channel, does not grow 
when adding qubits to the quantum register, as 
qubits can move simultaneously along their own channel. Therefore,
$\tau_{{\rm ch}}$ is not usually taken into account when determining the typical 
times scales of quantum algorithms. 
However, the fact that each qubit have its own quantum channel  
throughout the whole circuit, unless some measurement intervene 
to interrupt the quantum processing, is of absolute relevance, as it 
contributes to give quantum computation one of its most relevant 
features, namely \textit{reversibility}.

\subsection{Single-qubit gates}
\label{s.singlequbitgates}
Let us now consider single-qubit gates G1, usually pictured as 
boxes along one line: 
they represent operations on one 
single qubit that are assumed unitary, so as to guarantee that
the norm of the input qubit state be conserved.
\begin{figure}[H]
		\begin{center}
		\begin{tikzpicture}
\draw (-2,-3) node {$\ket{Q}_{\rm in}$};
\draw (4,-3) node {$\ket{Q}_{\rm out}$};
\draw (0,-2) rectangle (2,-4)[line width=1];
\draw (-1,-3) -- (0,-3) [line width=2];
\draw (2,-3) -- (3,-3) [-, line width=2];
\draw (1,-3) node {G1};
\draw (1,-4.5) node {$\tau_{{\rm G1}}$};
		\end{tikzpicture}
\caption{One-qubit gate; $\tau_{{\rm G1}}$ is the time that the gate 
takes to accomplish} its task.
\end{center}
\end{figure}
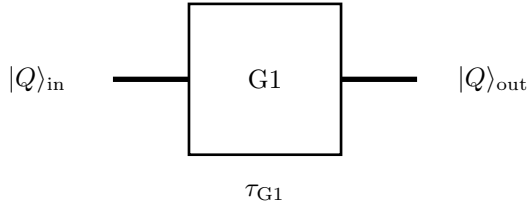
From a physical perspective, referring to the postulates of 
quantum mechanics, this entails that G1 represents the dynamical
evolution of the qubit, and can therefore be realized by properly 
designing an Hamiltonian \eqref{e.HQ}, according to
\begin{equation}
{\rm G1}\ket{Q}_{\rm in}=e^{-iH{_{_Q}}\tau_{_{\rm G1}}}\ket{Q}_{\rm in}=\ket{Q}_{\rm out}~,
\label{e.GHQ}
\end{equation}
where $\tau_{{\rm G1}}$ is the time which is physically necessary for the logical 
gate to complete its action. In order to make this point clearer, let us 
explicitely consider one of the prototypical single-qubit gate, namely 
the gate $X\rightarrow\left(\begin{array}{cc}
0 & 1 \\
1 & 0 \\ \end{array}\right),$ performing the spin-flip and corresponding 
to the Pauli operator $ \sigma^x $. Realizing a gate $X$ means to 
determine the Hamiltonian $H_{{Q}}$ and the time $\tau_{X}$ such 
that Eq.~\eqref{e.GHQ} holds, with G1=$X$. To this purpose, we notice 
that
\begin{equation}
e^{-i\sigma^xt}=\sum_n \frac{(-it)^n}{n!}(\sigma^x)^n=
\end{equation}
\begin{equation}\label{sin,cos}
=\left[\sum_{n}\frac{-i(t)^{2n+1}}{(2n+1)!}\right](\sigma^x)^{2n+1} 
+ \left[\sum_{n}\frac{(-it)^{2n}}{(2n)!}\right]\mathbbm{1}
= i\sin t \sigma^x-\cos t \mathbbm{1}~,
\end{equation}
implying $iX=\exp(-i\sigma^x \frac{\pi}{2})$: the action of a spin-flip 
gate is thus seen to be obtained by acting on $Q$ with the Hamiltonian 
$H_{{Q}}=g\sigma^x$ for the time interval 
\begin{equation}
\tau_{{X}}=\frac{\pi}{2g}~, 
\label{e.tauX}
\end{equation}
where $g$ is a generic coupling. 
When the qubit is represented by the spin degree of freedom of a 
particle with $S=1/2$, the above realization is readily obtained by 
applying a magnetic field along the $x$ direction for a finite time 
interval $\tau_{X}$, in which case 
$g=\gamma\mu_{_{\rm B}}h$ with $\gamma$ the gyromagnetic ratio of the 
particle, $\mu_{_{\rm B}}$ the Bohr magneton, and $h$ the intensity of 
the field. 
This example clarifies in what sense $\tau_{{\rm G1}}$ is not just 
a label, but must be regarded as a genuine physical time, depending on 
fundamental constants and tunable Hamiltonian parameters. \\
\subsection{Two-qubit gates} 
\label{s.twoqubitgates} 
Two-qubit gates G2,  are most often indicated as boxes across pairs 
of lines, 
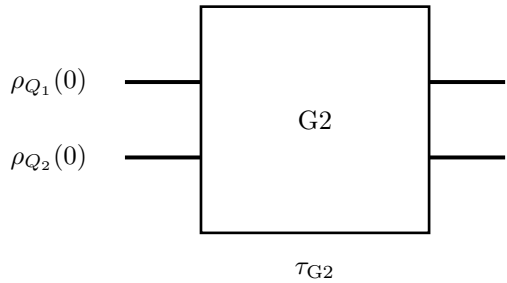
\begin{figure}[H]
\begin{center}
\begin{tikzpicture} 
\draw (7,0.5) -- (8,0.5)[line width=1.5];
\draw (8,1.5) rectangle (11,-1.5)[line width=1];
\draw [line width=1.5] (11,0.5) -- (12,0.5);
\draw (7,-0.5) -- (8,-0.5) [line width=1.5];
\draw [line width=1.5] (11,-0.5) -- (12,-0.5);
\draw (9.5, 0) node {G2};
\draw (6,0.5) node {$ \rho_{Q_1}(0) $};
\draw (6,-0.5) node {$ \rho_{Q_2}(0) $};
\draw (9.5,-2) node {$\tau_{{\rm G2}}$};
\end{tikzpicture}
\end{center}
\caption{Two-qubit gate; $\tau_{\rm G2}$ is the time that the gate takes 
to accomplish its task.} 
\end{figure}
\noindent thus visualizing that their  
action is essentially non-local. 
The fact that whatever G2 has input and output consisting
of the same quantum register implies that any such gate represents 
a unitary dynamics of the register itsef, i.e. an operator acting on 
${\cal H}_{QR}={\cal H}_{Q_1}\otimes{\cal H}_{Q_2}$ such that
\begin{equation}
{\rm G2}\ket{QR}_{\rm in}=e^{-iH_{_{QR}}\tau_{_{\rm G2}}}\ket{QR}_{\rm in}=
\ket{QR}_{\rm out}~,
\label{e.G2}
\end{equation}
where $\tau_{{\rm G2}}$ is the time which is physically necessary for the 
gate to complete its action, and the hermitian operator $H_{QR}$ is some
Hamiltonian for the quantum register.
Despite their identical formal description in so far as one considers 
$QR$ as a whole quantum system, G1 and G2 are profoundly different 
regarding their action on the two qubits $Q_1$ and $Q_2$, separately.
In fact, two-qubit gates must be intended as entangling gates, which 
implies that $H_{QR}$ embodies an effective interaction 
between $Q_1$ and $Q_2$ that prevents G2 to be written as tensor 
product of single-qubit gates. 
Furthermore, the dynamical entanglement generation implied by G2 makes
the single-qubit evolution non-unitary, a fact that lies at the hearth 
of the high efficiency of quantum computation, but also of 
the rather puzzling ban on the description of what happens to each qubit 
while being processed by a G2, as extensively discussed in 
Sec.~\ref{s.non-markovian}.

\subsection{Measurement apparatuses}
\label{s.measurementapparatuses}
Measurement apparatuses are usually indicated by a "meter"-symbol (see 
(4) in Fig.\ref{f.qalgorithm}) that 
put a final end to the quantum channel upon which it lies
Sometimes a classical-communication channel (usually represented by a 
double line) emerges from the symbol, indicating that 
the measurement process is the way we extract classical information out 
of quantum systems. This graphical representation recaps all the 
foundational issues that make the measurement {\it process} to actually 
be a measurement {\it problem}, whose discussion goes beyond the scope 
of this work. However, it is important to mention that measuring {\it is} a 
dynamical process, embodying entanglement generation between the 
observed quantum system and the apparatus itself, and therefore 
implying non-unitary evolution for both subsystems; in fact, despite the 
initial entanglement generation can be essentially described as in the 
G2 case, the fact that the apparatus is macroscopic, which is a 
necessary condition if it has to produce a signal that can be detected 
and characterized by our reading instruments, makes the description of 
the measuring finale a much debated matter 
(see for instance Refs.\cite{Schloss07,WheelerZ83}). 
Without commenting further upon this point, we just notice that
most of the features of OQS dynamics that we will discuss in the next 
section do actually characterizes also the measurement process.

\section{Non-unitary dynamics of open quantum systems}
\label{s.entanglement}

\begin{figure}[H] 
\begin{center}
\begin{tikzpicture}
\draw [->, line width=2] (-0.5,3) -- (4,3);
\draw (-3.9,3) node {$\ket{Q_1(0)}\bra{Q_1(0)}\otimes\ket{Q_2(0)}\bra{Q_2(0)}$};
\draw (5.6,3) node {$\ket{QR(t)}\bra{QR(t)}$};
\draw [->, line width=2] (4.5,2.5) -- (4.5,-1);
\draw (1.5,3.5) node {$ U_t^{(QR)}$};
\draw (5.3,1) node {$ {\rm Tr}_{_{Q_2}}$};
\draw (5,-1.5) node {$\rho_{_{Q_1}}(t)$};
\draw [->, line width=2] (-1.5,2.5) -- (-1.5,-1);		
\draw (-2.6,-1.5) node {$\ket{Q_1(0)}\bra{Q_1(0)}$};
\draw (-2.3,1)node {${\rm Tr}_{_{Q_2}}$};
\draw [->, line width=2, color=red] (-0.5,-1.5) -- (4,-1.5);
\draw (1.5,-2) node {$ \mathcal{E}_{t;\ket{Q_2(0)}}$};
\end{tikzpicture}
\end{center}
\caption{Global and local evolution of $QR$ and $Q_1$.}
\label{f.figUDM}
\end{figure}
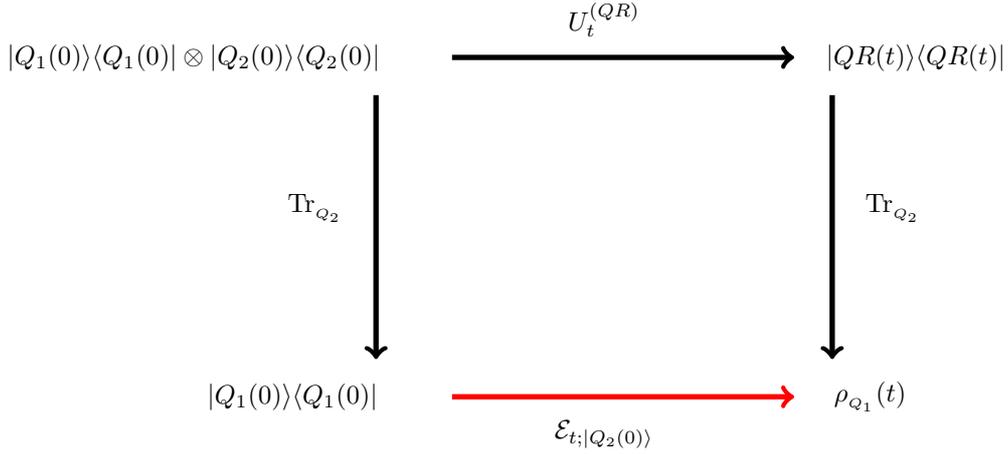

In order to better understand the peculiarities of the dynamical 
process underlying the G2 functioning, let us consider the following:
Take a two-qubit quantum register, initially prepared in the state 
$\ket{QR(0)}$, and let it evolve under a unitary evolution 
\eqref{e.QRevolution}. Adopting a local viewpoint on, say, $Q_1$, the 
only tool we can use to study its behaviour at any $t>0$ is 
\begin{equation}
\rho_{Q1}(t)={\rm Tr}_{{Q_2}}\ket{QR(t)}\bra{QR(t)}~,
\label{e.rhoQ1}
\end{equation}
and it therefore make sense to define and characterize a possible map
\begin{equation}
{\cal M}: \rho_{Q_1}(0)\longrightarrow\rho_{Q_1}(t)
\label{e.map}
\end{equation}
in terms of its dependencies and properties.
As for the former, while a dependence on $U_t^{(QR)}$ is expected and 
recognized as harmless, one cannot allow ${\cal M}$ to depend on 
$\rho_{Q_1}(0)$ as this would make it essentially ill-defined
(a superoperator cannot depend on the operator upon which it acts!).
In fact, using Eq.\eqref{e.QRevolution} for expressing
Eq.~\eqref{e.rhoQ1}, one can show that Eq.~\eqref{e.map} 
defines a proper superoperator if and only if 
$\ket{QR(0)}=\ket{Q_1(0)}\otimes\ket{Q_2(0)}$:
when this is the case ${\cal M}$ is dubbed universal 
dynamical map (UDM) and usually indicated by ${\cal 
E}_{t;\ket{Q_2(0)}}$. The dependence on $\ket{Q_2(0)}$ is  most 
often understood but, due to its relevance as far 
as our considerations are concerned, we prefer to keep it evident.
A UDM can be used to establish the precise relation 
between global and local evolution, of $QR$ and $Q_1$ respectively, 
as represented in Fig.~\ref{f.figUDM}.
Notice that what makes the above scheme fully consistent is
the possibility of drawing the line, marked in red on purpose, 
describing the action of the UDM ${\cal E}_{t;\ket{Q_2(0)}}$;
this exemplifies the more general result that, unless the 
initial state of the global system is separable, there is no way to 
formally describe the evolution of its subsystems by superoperators 
acting on their respective reduced density operators.
This result is of paramount relevance, as further discussed in the next 
Section.

\section{Non-markovian evolution of entangled systems}
\label{s.non-markovian}

Consider a two-qubit quantum register with a given hamiltonian that 
makes its initially separable state, $\ket{Q_1(0)}\otimes\ket{Q_2(0)}$,
to unitarily evolve in time. The corresponding $\rho_{Q_1}(t)$ and 
$\rho_{Q_1}(T)$  at a later time $T>t$ are both expressed by the same 
UDM, being equal to ${\cal E}_{t;\ket{Q_2(0)}}[\rho_{Q_1}(0)]$ and  
${\cal E}_{T;\ket{Q_2(0)}}[\rho_{Q_1}(0)]$, respectively.
However, we generally expect that the state $\ket{QR(t)}$ be entangled,
which means, according to the discussion of the above section, 
that the evolution of $\rho_{Q_1}$ from whatever $\tau>t$ on
cannot be described by a UDM, let alone ${\cal E}_\tau$.
\begin{figure}[H] 
\begin{center}
\begin{tikzpicture}
\draw [->, line width=0.5] (-1,0) -- (9,0);
\draw (9.5,0.03) node {\tiny\rm time};
\draw (0,0) circle (0.075)[fill=black];
\draw (4,0) circle (0.075)[fill=black];
\draw (8,0) circle (0.075)[fill=black];
\draw (-0.5,-0.5) node {$0$};
\draw (4,-0.5) node {$t$};
\draw (8.5,-0.5) node {$T$};
\draw (4,2.7) node {${\cal E}_T$};
\draw (2,0.7) node {${\cal E}_t$};
\draw (6,-1.5) node {{\large\bf ?}};
\draw [->, line width=1.5] (0,0) arc (145:40:4.9);
\draw [->, line width=1.5] (0,0) arc (145:40:2.4);
\draw [->, dashed, line width=1.5, color=red] (4,0) arc (220:310:2.6);
\end{tikzpicture}
\end{center}
\caption{Graphical rapresentation of non-markovian dynamics of open quantum 
systems.}
\label{f.nonmarkovian}
\end{figure}
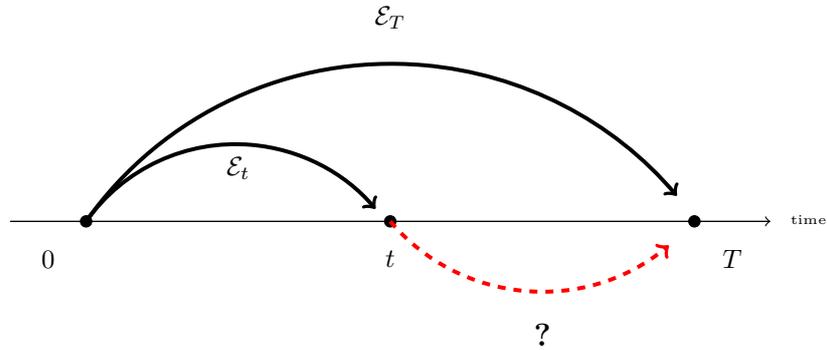
This implies that the composition law
\begin{equation}
{\cal E}_T={\cal E}_{(T-t)}{\cal E}_t
\label{e.composition}
\end{equation}
not only does not hold but it does not actually make any sense, as 
${\cal E}_{(T-t)}$ is not defined.
On the other hand, the validity of Eq.~\eqref{e.composition} defines a 
"markovian" quantum evolution. Therefore, if the global 
$QR$ evolves in such a way that entanglement is dynamically generated 
between $Q_1$ and $Q_2$, then both of them undergo a non-markovian 
dynamics. The entirity of the above argument makes it hold for whatever 
open quantum system, which justifies the recent outbreak of topics related 
with non-markovianity and its profound link with entanglement and its 
dynamical generation.

Let us however go back to the simplest case of OQS dynamics, i.e. that 
of $Q_1$ and $Q_2$ processed by the logical gate G2.
The above analysis, once related to the description given in 
Sec. \ref{s.entanglement}, tells us that we cannot actually ask what 
happens to, say, $Q1$ at some $t<\tau_{\rm G2}$, i.e. before the gate has completed its task, as this would drastically change its 
very same later evolution, and hence the G2 functioning itself.
Notice that this should not be intended as a ban on carrying out an 
actual measurement process without altering the state of $Q_1$ at a 
certain time $t$, but rather as a more profound lack of any prescription 
allowing one to determine how $Q_1$ will evolve 
from $t$ to $t+dt$, no matter how 
small $dt$ is, whenever $\ket{QR(t)}$ is an entangled state.
More formally, with reference to Eq.\eqref{e.G2}, no matter whether 
$H^{(QR)}$, $\tau_{\rm G2}$, and
$\ket{QR}_{\rm in}=\ket{Q_1(0)}\otimes\ket{Q_2(0)}$ are all well defined, 
yet there is no UDM that can possibly map $\rho_{Q_1}(t)$ into 
$\rho_{Q_1}(t+dt)$, if $Q_1$ and $Q_2$ are entangled at time $t$.

We cannot help noticing that two-qubit logical gates are the only 
elements of quantum algorithms, kept aside the final reading stage 
realized by quantum measurements, embodying dynamical entanglement 
generation, and hence affected by non-markovianity.

On the other hand, we do also know that quantum computation without 
two-qubit gates would not be different from classical 
computation in terms of computational speed \cite{Jozsa97,EkertJ98}, 
as the substantial difference in the way the processing time scales with 
respect to the length of the input register is essentially due to  
the modest power-law increase in the number of the required two-qubit 
gates (see for instance the gates-counting for the quantum fourier 
transform in Ref.\cite{NielsenC00} ). In brief: quantum 
parallelism realizes via two-qubit logical gates (as seen for instance 
in the Deutsch algorithm\cite{NielsenC00}).

The above two paragraphs suggests that what is somehow perceived as a 
detrimental feature of OQS dynamics, namely non-markovianity, lies 
indeed
at the hearth of the extraordinary efficacy of quantum processing.
In fact, surveying synonyms usually adopted in the OQS jargon give us a 
clue: "markovian" is used to mean "without memory" and, conversely,
"non-markovian" implies "with memory".

\section{Memory and effective processing}
\label{s.memory}
Although an extensive discussion of what is actually meant by markovian 
dynamics, in general and in the specific case of quantum evolution, 
goes far beyond the scope of this paper, we need mentioning that 
the general invalidity of Eq.\eqref{e.composition} has consequences as 
dramatic as that of preventing the time derivative of $\rho_{Q_1}$ to be 
a local function of time. In other terms, if we were to write an 
equation for $\rho_{Q_1}(t)$ this would generally result in an 
integro-differential equation \cite{Rivas12} 
\begin{equation}
\frac{d\rho_{\rm Q_1}(t)}{dt}=\int_0^t\,{\cal K}(t,u)\rho_{\rm Q_1}(u)\, 
du~,
\label{e.integrodiff}
\end{equation}
where the appearance of the integral formally represent the argument 
discussed in the previous section: in order to determine 
$\rho_{Q_1}(t+dt)$ the mere knowledge of $\rho_{Q_1}(t)$ is not 
sufficient, and the whole evolution from $\rho_{Q_1}(0)$ to 
$\rho_{Q_1}(t+dt)$ must indeed be considered.
Referring to the actual case of the evolution induced by a two-qubit 
gate, it is as if $\tau_{\rm G2}$ set a rhythm for the overall dynamics, 
in that it makes any shorter unit essentially meaningless.

Despite a mathematical analysis be extremely complex,
all of the above summons "memory" up. 
Indeed, the kernel ${\cal K}(t,u)$ in Eq.~\eqref{e.integrodiff} is 
usually referred to as "memory kernel", witnessing that 
non-markovian dynamics keeps memory of the previous evolution; on the 
contrary, markovian dynamics is classically defined as an evolution with 
no memory. On a qualitative basis, we think it 
does actually make sense to ascribe the superior power of quantum 
computation in processing information to the essentially un-forgetful 
evolution undergone by each single qubit passing through a two-qubit 
logical gate. Notice that the concept of an evolution with "memory" 
emerges due to the 
dynamical entanglement generation, as if it were the dynamical counterpart
of quantum non-locality.

Transferring arguments that have been developed by the 
quantum formalism to more general frameworks is often a risky procedure;
however, 
the idea that memory plays an essential role in the way 
information is processed is not at all abstruse. In fact, one of the 
most common experience when confronting with very rich informational 
contents, such as a piece of music, a 
poetry, or a reinassance masterpiece, is that no meaningful elaboration
would be possible were we unable to retain inputs during the 
whole listening, reading, or contemplation.

The above statement is so generic to border on irrelevance. If one 
considers, for instance, a piece of music, it is an obvious statement 
that our opinion on it, as well as the reaction it 
produces upon ourselves, emerges from the whole path that brings us 
from the silence before the first note to the silence after the very last 
one. There is no meaning in trying to get a sense out of one limited part 
of an artistic production. However, we believe it is possible to give a 
more formal content to the qualitative idea that memory has a role in 
perceiving any artistic suggestion, and we here sketch some related 
thoughts, with specific reference to musical works.  

Connections between quantum information and musical semantics have 
already been analyzed \cite{DallaChiara12}, suggesting 
that the tools of OQS dynamics might help understanding our capability of 
transforming impressive amounts of data into meaningful contents and 
experiences. In fact, one of the most striking connection between quantum 
computation and music consists of the graphic similarities between a 
quantum circuit (QC) and a musical score (MS): in both structures there 
are two orthogonal (from both the geometrical and the semantic viewpoint) 
directions along which events occur:
moving vertically we encounter different qubits in a QC and different 
instruments/voices in a MS, while the horizontal line corresponds to the 
time's flow in both cases.

The way scores are usually analyzed in music-theory textbooks is 
that of separately considering these two directions, so that melody
(horizontal line) and harmony (vertical line) become subject to separate 
analysis, carried on with different tools, such as the definition 
of themes and counter-themes rather then the characterization of chords 
over bass-lines, respectively.
In fact, there are rules that somehow establish a connection between 
these two directions: for instance, we are told that the seventh in a 
seventh chord (harmony) should resolve, i.e. move (melody), 
down. Also, we learn that one should not let the two notes of a fifth 
interval (harmony) to move (melody) so as to 
reproduce a fifth interval again (the so called "forbidden parallel 
motion" for the fifths). These guidelines somehow establish how the 
music should evolve at time $t+dt$, based on what has happened at time 
$t$, although considering some persistent notions such as the key and/or 
its major or minor character. 
The power of such rules in determining what will happen next is not as 
strong as that of a differential equation; however, we cannot help 
noticing that the kind of analysis they emerge from is one where 
short-term memory is given the same essential role it has in markovian 
dynamics. On the other hand, we all know that a musical discourse is 
everything but a deterministic step-after-step dynamical process, 
insofar as it is not even perceived in the same way by different listeners.

Inspired by these observations, we remind that also QC cannot be 
described as markovian evolution, specifically due to the presence of 
entangling gates, where entanglement is generated and information is 
processed in an essentially synthetic way. Indeed, the graphical symbols 
for these gates, as seen in Fig.\ref{f.qalgorithm}, are genuinely 
two-dimensional, in that they have both a finite length 
($\tau_{{\rm G}2}$) and a finite height (the number of qubits they 
process). Pushing the analogy forward, we think that it might 
be possible to develop an original way of reading a 
musical score based on the identification of two-dimensional structures, 
whose role in determining the evolution of the score is not visible 
unless one takes a truly global view on them.
We also like to think that the actual length and height of these 
structures be not objectively determined, so that different listeners
might partake in the discourse in different ways, depending on their 
personal attitude and/or capability of retaining perception of
the harmonic structure as time goes by.

\section{Aknowledgements}

We thank prof. Maria Luisa Dalla Chiara and prof. Elena Castellani 
for fruitful discussions and support.
This work is done in the framework of the Convenzione operativa 
between the Institute for Complex Systems of the italian National 
Research Council, and the Physics and Astronomy Department of the 
Univeristy of Florence.


\end{document}